# Data libraries as a collaborative tool across Monte Carlo codes

Mauro AUGELLI[1], Marcia BEGALLI[2], Mincheol HAN[3], Steffen HAUF[4], Chan-Hyeung KIM[3], Markus KUSTER[4], Maria Grazia PIA[5*], Lina QUINTIERI[6], Paolo SARACCO[5], Hee SEO[3], Manju SUDHAKAR[5], Georg WEIDENSPOINTNER[7], Andreas ZOGLAUER[8]

[1] *CNES, 31401 Toulouse, France*
[2] *State University Rio de Janeiro, 20550-013 Rio de Janeiro, Brazil*
[3] *Hanyang University, 133-791 Seoul, Korea*
[4] *Technische Universität Darmstadt, IKP, Germany*
[5] *INFN Sezione di Genova, 16146 Genova, Italy*
[6] *INFN Laboratori Nazionali di Frascati, 00044 Frascati, Italy*
[7] *MPE and MPI Halbleiterlabor, 81739 München, Germany*
[8] *University of California at Berkeley, 94720 Berkeley, CA, USA*

The role of data libraries in Monte Carlo simulation is discussed. A number of data libraries currently in preparation are reviewed; their data are critically examined with respect to the state-of-the-art in the respective fields. Extensive tests with respect to experimental data have been performed for the validation of their content.

## I. Introduction

Data libraries, consisting of tabulations of physics quantities originating from experimental or theoretical sources, are a widely used, essential tool in Monte Carlo simulation.

Their main purpose is to facilitate the simulation of physics processes by providing evaluated compilations of experimental data (or fits to them), or tabulations of theoretical quantities, which would be cumbersome to perform in the course of the simulation execution.

Data libraries could also play another valuable role in the context of Monte Carlo simulation application and benchmarking. Tabulations of fundamental physics quantities (cross sections, secondary particle spectra etc.) used by a Monte Carlo code could be a means for evaluating the effects of different physics modeling approaches in the same simulation environment of particle transport, geometry, user-defined cuts etc. The public provision of such tabulations should be promoted in the scientific community to facilitate the evaluation of possible systematic effects in simulation application results and to contribute to simulation validation efforts.

Recent activity concerning the creation of new data libraries meant for public distribution is reviewed in the following sections. Recent physics validation results, which question the accuracy of currently available data tabulations, are also discussed and suggestions for their update are proposed to better reflect the state-of-the-art in the associated field.

*Corresponding Author, MariaGrazia.Pia@ge.infn.it.

## II. Proton and α ionization data library

Recent progress in PIXE simulation[1] with Geant4[2,3] involved the development of simulation tools for the calculation of ionization cross sections by proton and α particle impact based on a variety of theoretical and empirical approaches: the ECPSSR[4] model and its variants (with Hartree-Slater corrections[5], with the "united atom" approximation[6] and specialized for high energies[7]), theoretical plane wave Born approximation, empirical models based on fits to experimental data collected by Paul and Sacher[8], Paul and Bolik[9], Kahoul et al.[10], Miyagawa et al.[11], Orlic et al.[12] and Sow et al.[13]. The empirical models concern K shell ionization by protons and α particles, and L shell ionization by protons; the theoretical models concern K, L and M shell ionization by protons and α particles.

The tabulated cross sections have been subject to extensive validation with respect to compilations of experimental measurements[8,14,15]. The comparison process involved rigorous statistical methods[16,17] to estimate the compatibility between the tabulated cross sections and experiment, and to evaluate the relative accuracy among the various modeling options. The full set of validation results is documented in a dedicated paper[1].

The above mentioned cross sections have been tabulated in the energy range between and 10 keV and 10 GeV; the tabulations of cross sections deriving from empirical models are limited to the energy range covered by the models themselves. The tabulations have been assembled in a data library, which is complemented by an example of basic software for retrieving the data and printing them. An example of the content of this data library is shown in **Figure 1**.

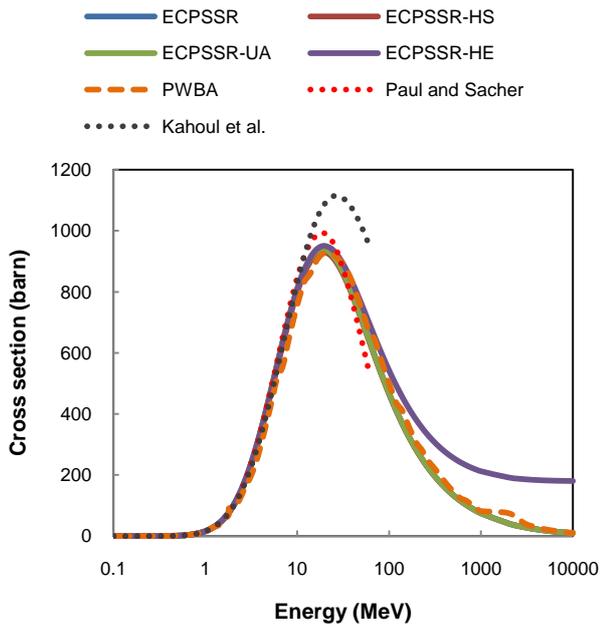

**Fig. 1** Cross sections for copper K shell ionization by proton impact included in the data library.

The data library is intended for public distribution by RSICC (the Radiation Safety Information Computational Center at Oak Ridge National Accelerator Laboratory). The final assembly for public release is being finalized at the time of writing this paper for the Monte Carlo 2010 Conference.

## III. Electron ionization data library

Recent progress in the simulation of low energy electron ionization[18] with Geant4 involved the development of simulation tools for the calculation of ionization cross sections by electron impact based on the Binary-Encounter-Bethe[19] model and the Deutsch-Märk[20] model.

The implemented cross sections, along with the cross sections tabulated in the Evaluated Electron Data Library[21] (EEDL), have been subject to extensive validation with respect to a large collection of experimental measurements, including more than 100 individual data sets and concerning more than 50 target elements. The comparison process involved statistical methods[16)17)] to estimate the compatibility between the calculated cross sections and experiment, and to evaluate the relative accuracy of the three modeling options. The full set of validation results is documented in a dedicated paper, which by far exceeds the limited page allocation of the papers presented at this conference; a sample of results is presented in another paper of this conference.

The above mentioned cross sections have been tabulated in the energy range between and 1 eV and 100 keV for elements with atomic number between 1 and 92 inclusive. The tabulations have been assembled in a data library, which is complemented by an example of basic software for retrieving the data and printing them. An example of the content of the data library is shown in **Figure 2**, along with cross sections contained in the EEDL data library and deriving from calculations[22] also used in Penelope.

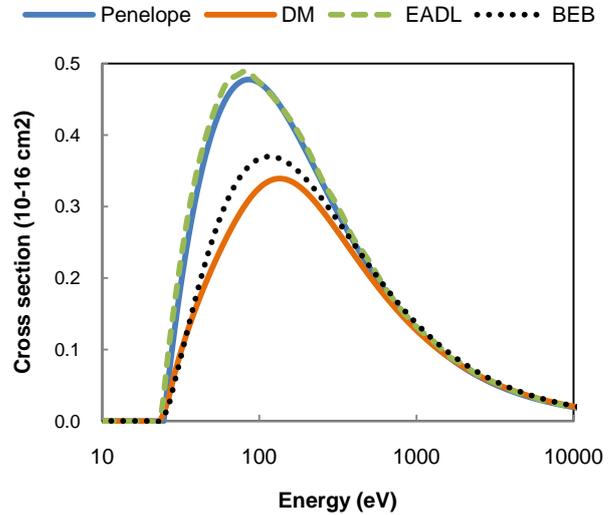

**Fig. 2** Electron impact ionization cross sections for Helium as a function of energy; the BEB and DM cross sections contained in the data library are shown together with the cross sections tabulated in the EADL library and deriving from calculations used in the Penelope code.

The data library is intended for public distribution; it will be released following the publication of the paper describing the new software developments and their validation. Since the validation process identified the new models as more accurate than EEDL at reproducing experimental measurements in the energy range below 250 eV, this data library could be a valuable alternative to EEDL in the lower energy range.

## IV. Improvement of EADL

Recent analyses[23)24)] for the experimental validation of parts of the Evaluated Data Library (EADL)[25] showed that some of its content does not reflect the state-of-the-art.

Regarding radiative transition probabilities, Hartree-Fock[26)27)] calculations appear more accurate[23] than the Hartree-Slater[28)29)] ones tabulated in EADL. Moreover, anomalies hinting to some accidental errors in the assembly of the library have been detected[30]; the flawed values are affected by errors amounting to orders of magnitude differences with respect to the original theoretical references[28)29)] from which they derive.

The inner shell binding energies of and ionization energies tabulated in EADL have also been subject to validation with respect to experimental data. A set of results from this analysis concerning the accuracy of inner shell binding energies is presented in another paper of this conference[24]. A sample of results concerning ionization energies is shown in **Figure 3**. The values in EADL appear in general less accurate than other compilations of electron binding energies available in the literature.

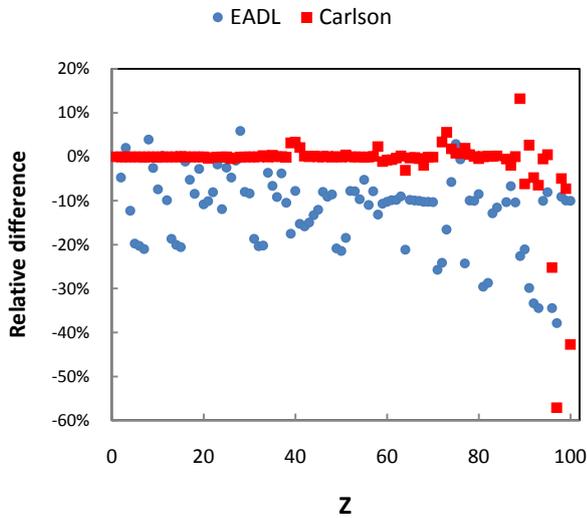

**Fig. 3** Comparison of ionization energies derived from EADL and Carlson's compilation with NIST reference values.

The full set of results deriving from this validation process is intended to be documented in a dedicated paper, whose publication will follow this conference.

These results suggest that an update of EADL would be beneficial to better reflect the state-of-the-art, which has evolved since the time of its last distributed version.

Along with the improvement of the content of EADL, a revision of its format to better match modern computational techniques would facilitate the used of this data library.

It should be stressed that EADL is an invaluable tool for Monte Carlo simulation, thanks to its wide collection of atomic parameters in one consistent environment; therefore it would be worthwhile to invest in its update.

**V. Review of radioactive decay data**

The quantitatively accurate representation of radioactive decays within a Monte Carlo simulation is of importance for a variety of applications such as dose calculations for medical, experimental and space flight purposes as well as estimating instrument responses in a wide range of experimental scenarios.

The current Geant4 radioactive decay simulation uses datasets which are based on the Evaluated Nuclear Structure Data Files (ENSDF)[31] to obtain half lives, decay branches, energy levels and level intensities of the decaying nucleus. It then passes the decayed nucleus to the photo-deexcitation process, which uses its own datasets to decay the nucleus into its ground state. The current database does not include references to the origins of the individual datasets or their actuality.

A comparison between the Geant4 datasets and the current version of the ENSDF shows disagreements in level energies for a considerable amount of isotopes as shown in **Figure 4**.

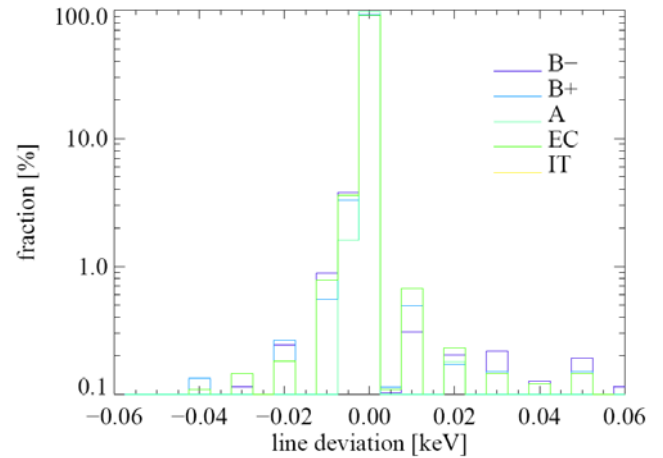

**Fig. 4** A comparison between the Geant4 datasets and the current version of the ENSDF.

There is currently no simple possibility for the end user to check the accuracy of the Geant4 data other than manually compare the entries to the state-of-the-art databases. An updated database which also includes references to the data origins would greatly facilitate this comparison. One should also consider providing decay data and deexcitation data as one consistent database, which would aid the identification of simulation inaccuracies in the common use case of a gamma ray detector detecting the deexcitation photons and not the levels of the decayed nucleus.

Finally an updated database could also make use of modern computational techniques such as an XML basis, which would facilitate parsing and structuring of the individual datasets.

**III. Conclusion**

Data libraries play an important role in Monte Carlo simulation.

New data libraries are currently in preparation, concerning ionization cross sections by electron, proton and α particle impact. Their accuracy has been estimated by means of extensive comparisons with experimental data.

Recent tests have shown to need of updating some parts of EADL to achieve better accuracy. Sources for such an improvement have been identified.

Provision of data libraries for open circulation within the scientific community should be promoted. Publicly available data libraries could facilitate comparisons of simulations based on different codes, as well as sharing physics modeling features in a variety of simulation environment.

The complete associated results are documented and discussed in depth in dedicated papers.

**Acknowledgment**

The authors express their gratitude to CERN for support to the research described in this paper.

The authors thank Sergio Bertolucci, Elisabetta Gargioni, Simone Giani, Berndt Grosswendt, Vladimir Grichine, Andreas Pfeiffer and Alessandro Zucchiatti for valuable discussions.

RSICC helpful collaboration for the assembly and distribution of data libraries is acknowledged.

to the research described in this paper.

The authors thank Sergio Bertolucci, Elisabetta Gargioni, Simone Giani, Berndt Grosswendt, Vladimir Grichine, Andreas Pfeiffer and Alessandro Zucchiatti for valuable discussions.

RSICC helpful collaboration for the assembly and distribution of data libraries is acknowledged.